\newcommand{\ale}{\ \raisebox{-.3ex}{$\stackrel{<}{\scriptstyle \sim}$}\ }
\newcommand{\age}{\ \raisebox{-.3ex}{$\stackrel{>}{\scriptstyle \sim}$}\ }

\documentstyle{mn}

\input psfig

\title[Warped discs and the direction of jets in AGN]
	{Warped discs and the directional stability of jets in \\
	Active Galactic Nuclei}
	
\author[P. Natarajan and P.J. Armitage]{Priyamvada Natarajan and Philip J. Armitage \\
	Canadian Institute for Theoretical Astrophysics, McLennan 
	Labs, 60 St George Street, Toronto, Ontario M5S 3H8, Canada}	
 	
\begin{document}

\maketitle

\begin{abstract}
Warped accretion discs in Active Galactic Nuclei (AGN) exert a torque
on the black hole that tends to align the rotation axis with the 
angular momentum of the outer disc. We compute the magnitude of this 
torque by solving numerically for the steady state shape of the warped disc, 
and verify that the analytic solution of Scheuer \& Feiler (1996) provides 
an excellent approximation. We generalise these results for discs with strong 
warps and arbitrary surface density profiles, and calculate the timescale 
on which the black hole becomes aligned with the angular momentum in the outer 
disc. For massive black holes and
accretion rates of the order of the Eddington limit the alignment timescale is 
always short ($\ale 10^6$ yr), so that jets accelerated from the inner disc 
region provide a prompt tracer of the angular momentum of gas at large radii 
in the disc. Longer timescales are predicted for low luminosity systems, 
depending on the degree of anisotropy in the disc's hydrodynamic 
response to shear and warp, and for the final decay of 
modest warps at large radii in the disc that are potentially observable 
via VLBI. We discuss the implications of this 
for the inferred accretion history of those Active Galactic Nuclei whose jet 
directions appear to be stable over long timescales. The large energy deposition 
rate at modest disc radii during rapid realignment episodes should make such 
objects transiently bright at optical and infrared wavelengths.
\end{abstract}

\begin{keywords}	
	accretion, accretion discs -- black hole physics -- galaxies: active --
	galaxies: jets -- galaxies: nuclei -- hydrodynamics
\end{keywords}

\section{Introduction}
The angular momentum of gas accreting onto massive black holes in Active 
Galactic Nuclei (AGN) is likely to change with time, either as a result 
of mergers funneling fresh gas towards the nucleus (Hernquist \& Mihos 1995), 
or because of radiation or disc wind driven warping instabilities in an 
existing accretion flow (Pringle 1996, 1997; Maloney, Begelman \& Pringle 1996;
Maloney, Begelman \& Nowak 1998).
In either case, the angular momentum of gas at large radius in the disc will 
be misaligned with the rotation axis of the black hole, while at small 
radius the combined action of viscosity and differential precession induced 
by the Lense-Thirring effect (Lense \& Thirring 1918) leads to alignment of 
the disc and hole angular momenta (Bardeen \& Petterson 1975; 
Kumar \& Pringle 1985). Achieving this 
alignment requires that the black hole exert a torque on the disc gas, and 
implies that an equal and opposite torque act on the black hole itself. Over 
time, this causes a change in the spin axis of the hole towards alignment 
with the large angular momentum reservoir provided by the disc at large radius. 

Whether the timescale for alignment is long or short compared to the lifetime
of an AGN has implications for our understanding of the accretion history 
of these objects. If the black hole is spinning, then jets, regardless of whether 
they derive power from the aligned inner disc or directly from the hole 
itself (Blandford \& Znajek 1977; see also Ghosh \& Abramowicz 1997;
Livio, Ogilvie \& Pringle 1999),
trace the spin of the black hole. If the 
alignment timescale is long, then we expect that the jet direction reflects 
the initial spin of the black hole established during (or prior) to the 
formation of the host galaxy (Rees 1978). Moreover the jet direction is 
expected to be stable over time, irrespective of variations in the angular
momentum of accreting gas. Conversely, if the timescale is short then
jets define the angular momentum of inflowing material, and the 
observation of systems where they appear to be stable over long 
time periods of $10^7$ - $10^8$ yr (Alexander \& Leahy 1987; 
Liu, Pooley \& Riley 1992; Scheuer 1995) 
implies a constancy in the average angular momentum of accreting gas.

The rate of realignment was computed by Scheuer \& Feiler (1996), who 
found an approximate analytic solution to the equations governing the 
evolution of a warp. The timescale depends on two poorly known 
viscosities, $\nu_1$ and $\nu_2$, which correspond to the $(R,\phi)$ 
and $(R,z)$ components of the shear as defined by Papaloizou \& Pringle (1983). 
If $\nu_1 \sim \nu_2$ (i.e. the timescale for warp evolution is comparable 
to that for the evolution of the surface density) then the 
formula of Scheuer \& Feiler (1996) is similar to that previously assumed by Rees (1978).
Subsequently, Natarajan \& Pringle (1998) revisited the problem, and pointed 
out that for the parameters of AGN discs the assumption that $\nu_1 \sim \nu_2$ 
is at variance with the results of detailed analyses of the hydrodynamics 
of viscous discs, which suggest instead that $\nu_2 \gg \nu_1$. Recalculating 
the alignment timescale with this modification they found that it was 
{\em short} --- much less than any reasonable estimate of AGN lifetime --- 
and used this to argue that the spin of black holes in AGN, and the direction 
of jets produced in the inner regions of the accretion flow, rapidly adjust 
to trace the angular momentum of gas at large radius.

The analysis in Natarajan \& Pringle (1998) employed the same torque
formula as Scheuer \& Feiler (1996), which was derived under the restrictive 
conditions of small amplitude warps and a disc viscosity that is constant 
with radius (or, equivalently, that in a steady-state the disc surface
density is independent of radius). Since AGN may well harbour strongly
warped discs, and almost certainly have more complex surface density 
profiles, our goal in this paper is to generalise their results. Section 2
below sets out the equations for the evolution of a warped disc, which 
we solve numerically in Section 3 for comparison with the approximate 
analytic solution. The resulting torque is calculated in Section 4, 
and the alignment timescale derived for a simple AGN disc model in 
Section 5. Section 6 discusses the implications of our results.

\section{Warped disc equations}

\subsection{Governing equations}

The evolution of a thin, warped accretion disc in 
a Keplerian potential was studied by Papaloizou \& Pringle (1983). For 
a disc with a surface density profile $\Sigma(R,t)$, with angular momentum 
at radius $R$ parallel to the unit vector ${\it \hat{l}}$, the evolution can 
be expressed most compactly in terms of an equation for the angular 
momentum density ${\bf L} = (GMR)^{1/2} \Sigma {\it \hat{l}}$. Adopting the 
simplest description (Pringle 1992; see also Papaloizou \& Pringle 1983), the 
time evolution is given by,
\begin{eqnarray}
{\frac{\partial {\bf L}}{\partial t}}\,&=&\,{\frac{3}{R}}\,{\frac{\partial}{\partial R}
}\,\left [{\frac{R^{1/2}}{\Sigma}}{\frac{\partial}{\partial R}}
(\nu_1\,\Sigma\,R^{1/2})\,{\bf L}\right ]\,\nonumber \\ 
&+&{\frac{1}{R}}\,{\frac{\partial}{\partial R}}\,\left 
[(\nu_2\,R^2\,{\left\vert{\frac{\partial {\it \hat l}}
{\partial R}}\right\vert^2}\,-\,{\frac{3}{2}}\,\nu_1)\,{\bf L} \right ]\nonumber \\ 
&+&{\frac{1}{R}}\,{\frac{\partial}{\partial R}}\,\left 
({\frac{1}{2}}\,\nu_2\,R\,|{\bf L}|\,{\frac{\partial \hat l}{\partial R}} \right )
\,+\,{{{\vec \omega_p} \times {\bf L}} \over R^3}.
\label{eq1}
\end{eqnarray} 
Here $\nu_1$ and $\nu_2$ are viscosities acting on the $(R,\phi)$ and 
$(R,z)$ components of the shear, as defined by Papaloizou \& Pringle (1983). 
Neither is equal to the shear viscosity `$\nu$' that enters into the 
Navier-Stokes equations, and the general relation between these three quantities 
is extremely complex (Ogilvie 1999). For the parameters of AGN discs, the 
simplest assumptions suggest that $\nu_2 \gg \nu_1$ (Natarajan \& Pringle 1998), 
though for most of this paper we will allow both viscosities to be general 
power laws with radius,
\begin{eqnarray} 
 \nu_1 & = & \nu_{10} R^\beta \nonumber \\ 
 \nu_2 & = & \nu_{20} R^\beta.
\label{eq2}
\end{eqnarray}
For thin, vertically averaged accretion disc models, the shear viscosity $\nu$ 
is a power-law in radius in a given opacity regime (e.g. Frank, King \& Raine 1992). 
Later, the index $\beta$ will be chosen to correspond to the scaling expected 
at radii in the disc where the bulk of the realignment torque acts. However, 
for a general warped disc, the radial run of both $\nu_1$ and $\nu_2$ will also 
depend on the {\em shape} of the warp (Ogilvie 1999). This effect is not 
taken into account in this paper. Our assumption that $\nu_1$ and $\nu_2$ 
are simple power-laws in radius, independent of warp shape, is therefore 
strictly valid in the limit of small warps, and is an approximation for 
larger amplitude warps. 

The range of behaviour allowed by equation (\ref{eq1}) is considerably more 
diverse than simple diffusion of ${\bf L}$. Nonetheless, we will make 
frequent use of the characteristic timescales for diffusion of surface density 
and warp, which we define conventionally as,
\begin{equation} 
 t_{\nu_1} = { R^2 \over \nu_1}, \ \ t_{\nu_2} = { R^2 \over \nu_2}.
\label{eq2}
\end{equation} 
The final term on the right hand side of equation (\ref{eq1}) represents the effect 
of Lense-Thirring precession on the disc. For a black hole with mass $M$, we take 
${\vec \omega_p} = {\bf \omega_p} (0,0,1)$, where
\begin{equation} 
 {\bf \omega_p} = 2 a c \left( {GM} \over c^2 \right)^2
\label{eq3}
\end{equation}
where $a$ is the dimensionless angular momentum of the black hole. 

Several assumptions and simplifications are necessary to derive this 
equation, and these are set out in detail in Papaloizou \& Pringle (1983)
(see also Pringle 1999; Ogilvie 1999). Particularly important is the boundary
between the diffusive behaviour described by equation (\ref{eq1}) and wave-like
evolution, which occurs when the usual Shakura \& Sunyaev (1973) viscosity parameter
$\alpha \approx H / R$, where $H$ is the disc scale height (Papaloizou \& Lin 
1995). The observational constraints on $\alpha$ in AGN discs are weak 
(Siemiginowska \& Czerny (1989) and Siemiginowska, Czerny \& Kostyunin (1996) 
suggest $\alpha=10^{-1} - 10^{-2}$), but in the ionized parts of the disc 
it is plausible to assume that $\alpha \sim 0.1$, as in better studied 
disc systems (eg. Cannizzo 1993). In any case, theoretical estimates 
for $H/R$ at the relevant radii (Collin-Souffrin \& Dumont 1990) 
suggest that in AGN $\alpha \gg H/R$, so that we are 
safely in the diffusive regime. In this respect AGN discs are very
different from protostellar discs, for example, where wave-like behaviour 
is likely to be important (e.g. Larwood et al. 1996). We also 
note that although Papaloizou \& Pringle (1983) considered only 
small amplitude warps describable using linear perturbation theory, 
Ogilvie (1999) has shown that strongly warped discs can also 
be described using a similar approach.

\subsection{The Scheuer \& Feiler analytic solution}

Scheuer \& Feiler (1996) obtained an approximate steady state analytic solution to 
equation (\ref{eq1}) for the case where $\beta=0$, in the limit where the angle 
of misalignment of the disc and the hole spin was small. In this case, the 
surface density obeys the usual relation for a planar disc with accretion rate 
$\dot{M}$ and zero torque inner boundary condition imposed at $R_{\rm in}$,
\begin{equation} 
 \nu_1 \Sigma = { \dot{M} \over {3 \pi} } \left( 1 - \sqrt{R_{\rm in} \over R} \right).
\label{eq4}
\end{equation} 
The shape of the disc is given by ${\it \hat{l}} = (l_x,l_y,l_z)$, where
\begin{eqnarray}
 l_x = K e^{-\phi} \cos \phi \nonumber \\
 l_y = K e^{-\phi} \sin \phi
\label{eq5}
\end{eqnarray} 
and $\phi = 2 (\omega_p / \nu_2 R)^{1/2}$. The constant $K$ depends on the 
disc inclination $i$ at large radius, at large radius $K = \sin i$.
 
\subsection{Numerical solutions}

The above analytic solution applies only for $\beta = 0$, and as derived 
assumes that the disc warp is of small amplitude. Relaxing these assumptions 
requires a steady state numerical solution of equation (\ref{eq1}). 

For $\beta < 1/2$, steady state solutions can be calculated efficiently by 
solving directly the ordinary differential equations (ODEs) obtained by setting 
$\partial \Sigma / \partial t = 0$ in equation (\ref{eq1}). We give a full 
description of our methods in the Appendix, but briefly we have found it 
simplest to use equations expressed in terms of the surface density 
$\Sigma(R)$ and unit tilt vector ${\hat l} = {\bf L} / \vert {\bf L} \vert$, 
which can readily be solved iteratively using a finite-difference technique
described by Pereyra (1979). We use the Numerical Algorithms Group's
implementation, and impose boundary conditions of $\Sigma (R_{\rm in}) = 0$, 
$\Sigma (R_{\rm out}) = \Sigma_{\rm out}$, $\hat l (R=R_{\rm in}) = (0,0,1)$, 
$\hat l (R=R_{\rm out}) = \hat l_{\rm out}$. These boundary 
conditions {\em assume} that the inner disc is aligned with the spin 
axis of the hole, we therefore have to check post facto that this is indeed a 
valid assumption for each model calculated.

For $\beta \ge 1/2$, we have been unable to obtain a converged numerical 
solution to the steady state ODEs using these methods. In this regime, we
instead evolve the time dependent equation (\ref{eq1}) from an arbitrary 
initial condition (of a flat, uniformly tilted disc) until a steady state
is obtained, using the numerical method described by Pringle (1992). This 
is straightforward but computationally expensive, since many viscous times
at the outer edge of the disc are required to obtain a steady state 
solution. Defining the viscous time of the disc by $t_{\nu_1} = R_{\rm out}^2 / \nu_1$, 
we find that more than 10 viscous times of evolution are necessary. Moreover, 
for either method capturing the torque accurately requires a large range 
between the inner and outer disc radii, of at least 
$R_{\rm out} / R_{\rm in} \age 10^4$. For these reasons, we are only 
able to obtain solutions using the time dependent code for high values 
of $\beta$ ($\beta=1.5$), when the viscous time is a weak function 
of increasing radius.

\section{Steady-state disc shape}

\subsection{Comparison with the analytic solution}

\begin{figure}
 \psfig{figure=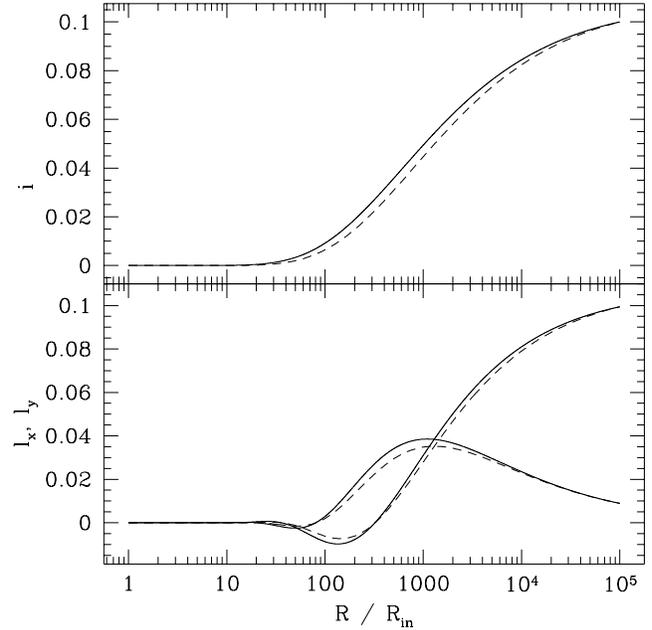,width=3.5truein,height=3.5truein}
 \caption{Comparison of the numerical solution to the full warped disc
          equation with the approximate solution derived by Scheuer \& 
          Feiler (1996). The parameters are $\beta=0$, $\nu_{10}=\nu_{20}=1$, 
          and $\omega_p = 200$. The disc at $R_{\rm out} = 10^5 R_{\rm in}$ 
          is fixed at an inclination angle of 0.1 with respect to the spin axis of 
          the black hole. The upper panel compares the inclination as a function 
          of radius in the exact solution (solid line) and the analytic solution 
          (dashed line). The lower panel shows the same comparison for two components
          of the tilt vector, $l_x$ and $l_y$. The agreement is at the level of a
          few percent.}
 \label{f1}
\end{figure}

Figure {\ref{f1}} shows the numerical steady-state solution for the 
$\beta = 0$ case, and compares it to the approximate analytic solution. 
We adopt units in which $R_{\rm in} = 1$, and take $\nu_{10} = \nu_{20} = 1$. 
We choose $\omega_p = 200$, which gives an inner aligned region 
extending out to around $10^2 \ R_{\rm in}$. This is roughly appropriate 
for plausible AGN parameters, though we defer until later consideration 
of detailed disc models. Convergence to the limiting inclination at 
infinity is slow, so it is necessary to impose the outer boundary 
condition of $i = 0.1$ at a large radius, $R = 10^5 \ R_{\rm in}$. 
At this radius $l_x$ and $l_y$ are chosen to facilitate comparison 
with the analytic solution of equation (\ref{eq5}).

Evidently, the analytic solution provides an excellent description 
of the disc shape. The inner disc is aligned to the rotation axis 
of the hole out to close to $10^2 \ R_{\rm in}$, and warps to 
around half the limiting inclination by $10^3 \ R_{\rm in}$. Moving 
inward from infinity, the warp is twisted by an angle of $\sim \pi$ 
before being flattened effectively into the aligned plane.

\subsection{Aligned radius}

\begin{figure}
 \psfig{figure=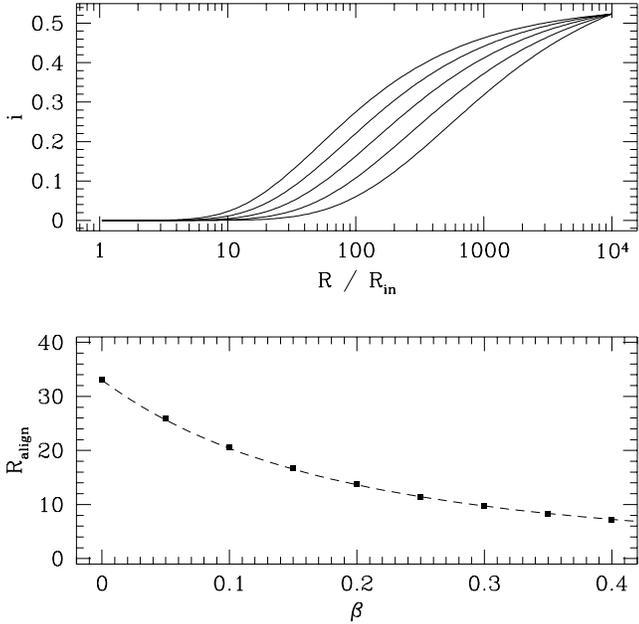,width=3.5truein,height=3.5truein}
 \caption{Disc shape for various values of $\beta$. The upper panel shows the 
          run of disc inclination $i$ for models with $\beta = 0$, 0.1, 0.2, 
          0.3 and 0.4 (larger $\beta$ models are warped to smaller $R$). All 
          models have $\omega_p = 200$ and $\nu_{10} = \nu_{20} = 1$, i.e.
          the viscous time for both surface density evolution and warp is 
          being kept fixed at the inner edge. The lower panel shows the 
          extent of the inner aligned region of the disc as a function of 
          $\beta$, where $R_{\rm align}$ is defined implicitly via 
          $i(R_{\rm align}) = \delta$, with $\delta = 0.01$. The dashed 
          curve shows the predicted relation if $R_{\rm align}$ scales with 
          the radius where the viscous timescale for the warp equals the 
          local precession timescale.}
 \label{f2}
\end{figure}

For an annulus in the disc, the local timescales for precession and for 
transmission of warp are given by,
\begin{equation} 
 t_p = {R^3 \over {\omega_p}}, \ \ 
 t_{\nu_2} = {R^2 \over {\nu_{20} R^\beta}}.
\label{eq8}
\end{equation} 
We expect that the disc will be aligned with the spin axis of the hole 
at radii where the precession timescale is much shorter than the timescale 
over which the disc can communicate warp inwards.
The characteristic radius of the aligned region $R_{\rm align}$ will then 
scale with the radius where $t_p = t_{\nu_2}$, which is given by,
\begin{equation} 
 R_{\rm align} = C_1 \left( {\omega_p} \over {\nu_{20}} \right)^{1/(1+\beta)},
\label{eq9}
\end{equation} 
where $C_1$ is a constant expected to be of the order of unity which we will use 
later to fit the numerical results. Note that this expression assumes both that the 
Lense-Thirring torque is strong enough to ensure that the inner disc {\em is} 
aligned, and that $\beta > -1$, so that $t_p$ increases with radius more rapidly 
than $t_{\nu_2}$.

Figure {\ref{f2}} shows how the disc shape varies with $\beta$ for a set of 
models with fixed $\omega_p$ and $\nu_1 = \nu_2 = 1$. From the numerical 
solutions, we compute $R_{\rm align}$ as the radius where $i$ first exceeds 
some small inclination $\delta$. With $C_1 = 0.165$ there is excellent 
agreement between the computed $R_{\rm align}$ and the scaling given by 
equation (\ref{eq9}), i.e. the simple timescale argument given above suffices
to give an accurate idea of the scaling of the radius out to which the hole can enforce 
inner disc alignment. 

\section{Torque}

The torque exerted on the misaligned disc as a result of the 
Lense-Thirring precession is given by the integral,
\begin{equation}
 - { {{\rm d}{\bf J}} \over {{\rm d}t}} = \int 
 { { {\vec \omega_p} \times {\bf L} } \over R^3 } 
 2 \pi R {\rm d} R.
 \label{eq6}
\end{equation} 
Of course an oppositely directed torque of the same magnitude is 
exerted on the black hole. Computing ${\dot{\bf J}}$ for the disc 
model shown in Fig.~{\ref{f1}} we find that the torque agrees with 
that obtained from the Scheuer \& Feiler solution at the level of a 
few percent (the small discrepancy being primarily due to the 
ignored inner boundary condition in the analytic solution, which 
leads to an error in the surface density in the warp region). The 
torque scales as $(\omega_p \nu_2)^{1/2} \dot{M}$, again as 
found analytically. We have computed additional models with 
large inclination angles at infinity (up to $80^\circ$ angle 
of misalignment) and for those models there is a larger 
discrepancy between the numerical and analytic results, but still 
only at the tens of percent level. 

\subsection{Radial dependence}

\begin{figure}
 \psfig{figure=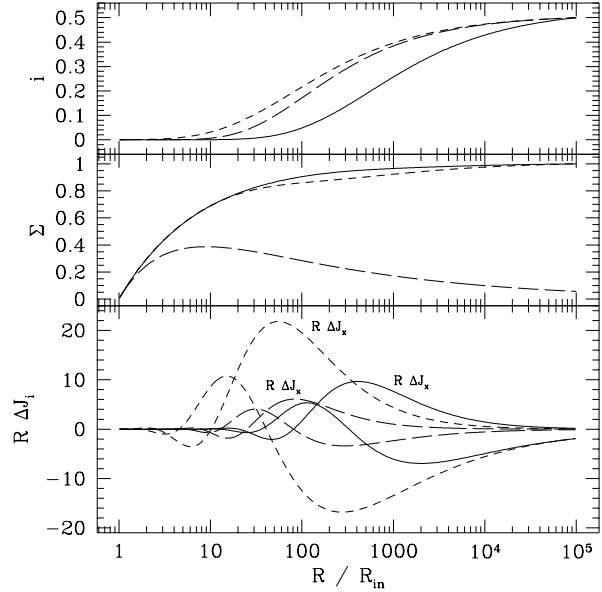,width=3.25truein,height=3.25truein}
 \caption{Radial dependence of the torque exerted on the disc. The upper 
          two panels show the inclination $i(R)$ and surface density 
	  $\Sigma(R)$ for three disc models, (i) $\nu_1 = \nu_2 = 1$ 
	  (solid line), (ii) $\nu_1 = 1$, $\nu_2 = 8$ (short dashed 
	  line) and (iii) $\nu_1 = \nu_2 = R^{1/4}$, all normalised 
	  to the same accretion and precession rate. The units are 
	  chosen such that $t_{\nu_1}$ and $t_{\nu_2}$ are also kept 
	  fixed at $R_{\rm in}$. The lower panel
	  shows where the resulting torques arise in the disc. We 
	  plot $R \Delta J_x$ and $R \Delta J_y$ for each model, 
	  where the total torque is given by the integral 
	  $\int \Delta J {\rm d}R$. The $R \Delta J_x$ components 
	  are labelled in the lowest panel of the figure.
	  For these parameters  
	  the greatest contribution arises at radii of $\sim 10^2 - 10^3 \ 
	  R_{\rm in}$. The surface density and torque units are arbitrary.} 
 \label{f3}
\end{figure}  

Figure {\ref{f3}} shows the tilt angle and surface density profile for 
three fiducial models, all normalised to the same accretion rate.
The solid line is computed for the parameters used for Figure {\ref{f1}}, 
namely $\beta=0$ and $\nu_{10} = \nu_{20}$. The short dashed line 
maintains $\beta=0$, but has $\nu_{10} = 1$, $\nu_{20} = 8$. Natarajan \& 
Pringle (1998) find that consideration of the hydrodynamics of a viscous 
disc suggests that in AGN, 
\begin{equation}
 { {\nu_2} \over {\nu_1} } \approx { 1 \over {2 \alpha^2} }
\label{eq_ratio}
\end{equation}
in which case this ratio of viscosities would be appropriate for 
a plausible Shakura-Sunyaev $\alpha$ of around 0.25. Finally 
the long dashed line illustrates the effect of a declining surface 
density profile, plotted is a model with $\beta = 1/4$ and $\nu_1 = \nu_2$.
All three of these models are chosen to have the same viscous timescale 
(both for the surface density and for warp) at the (arbitrary) choice 
of inner edge radius. In conventional thin disc models (Shakura \& Sunyaev 1973), 
fixing the viscous time at a given radius corresponds to a fixed disc 
thickness $H/R$ at that radius. We note that in such models 
the value of $\beta$ is fixed by solving for the vertical disc structure, i.e.
it is not a `parameter' that can be varied but rather a property of the disc. 
In Section 5 we discuss the appropriate value for $\beta$ in the region of the
disc where the dominant torque arises. Also shown for each of the models is 
where radially the greatest contribution to the torque arises, we plot the 
components of the integrand in equation (\ref{eq6}) (the extra factor of $R$ in 
the plotted quantity takes account of the logarithmic radial scale to give a true
impression of where the strongest torque on the disc is found).

For all the models, strong torques arise from a broad region that 
extends from just outside the alignment radius out to one or two 
orders of magnitude larger radius. The large radial range used 
in these calculations is thus essential to capture the torque 
accurately. Increasing the ratio of $\nu_2$ relative to $\nu_1$ 
reduces the timescale for diffusion of warp, $t_{\nu_2} = R^2 / \nu_2$, 
and allows the warp to push in closer to the black hole. At fixed accretion 
rate increasing $\nu_2$ even by this large factor leaves the surface density 
profile almost unaltered, and thus the net effect is simply to increase 
the magnitude of the torque and move it to smaller disc radius. Larger
values of $\beta$ likewise decrease the warp timescale at large radius and lead to 
a shrinkage of the aligned region. For the parameters used here, the disc 
shape is in fact rather similar for the two models in which $\nu_2 > 1$ 
at large radius. However, for a given accretion rate, models with higher $\beta$ 
have lower surface density at large radii, which tends to decrease ${\bf L}$ in 
equation (\ref{eq6}) and reduce the integrated torque. Both these effects 
have a significant impact on the calculation of the total torque.

\subsection{Scaling with $\beta$}

\begin{figure}
 \psfig{figure=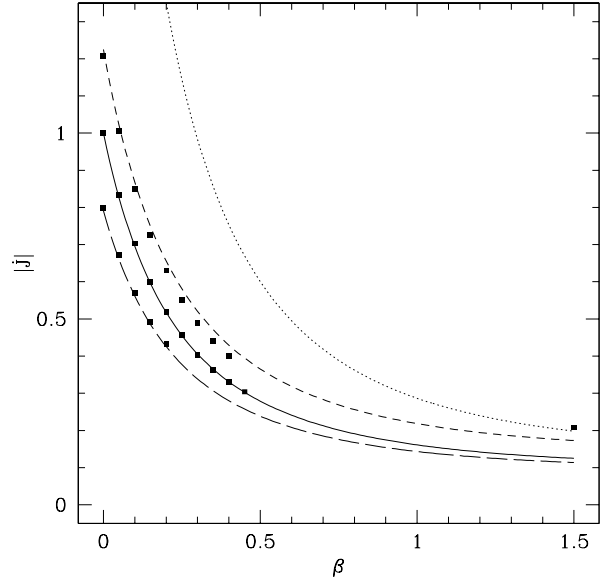,width=3.25truein,height=3.25truein}
 \caption{Scaling of the torque with $\beta$ at fixed mass accretion 
          rate, where the viscosity $\nu \propto R^\beta$.
          The points show numerical results, the curves the fitting 
	  formula described in the text. The main parameters are:
	  (i) solid line $\nu_{10}=\nu_{20}=1$, $\omega_p=200$, 
	  (ii) dotted line $\nu_{10}=\nu_{20}=0.1$, $\omega_p=200$, 
	  (iii) short dashed line $\nu_{10}=1$, $\nu_{20}=2$, $\omega_p=200$,
	  (iv) long dashed line $\nu_{10}=\nu_{20}=1$, $\omega_p=125$.
	  The torque on the hole drops by around an order of magnitude over 
	  the range of $\beta$ considered here.} 
 \label{f4}
\end{figure}

The numerical results show that the Scheuer \& Feiler (1996) solution, 
equation (\ref{eq5}), provides an accurate description of the disc shape 
and the resultant torque on the black hole. For this solution, the 
torque integral in equation (\ref{eq6}) has a magnitude 
given by,
\begin{equation} 
 \vert {\dot{\bf J}} \vert = { {\sqrt{2} K \dot{M}} \over {3 \nu_{10}}} 
 (GM)^{1/2} (\omega_p \nu_{20})^{1/2}.
\label{eq7}
\end{equation} 
Guided by the numerical results, a straightforward extension of this 
result to non-zero $\beta$ is possible. We first assume that the 
radius $R_w$ out to which we need to integrate equation (\ref{eq6}) to 
find the torque scales as does the aligned radius,
\begin{equation} 
 R_w = C_2 \left( {\omega_p} \over {\nu_{20}} \right)^{1/(1+\beta)},
\label{eq9x}
\end{equation} 
If we additionally assume that the disc shape is primarily a function of $R_w$, then
the form of the integral in equation (\ref{eq6}) suggests the following 
scaling with $\beta$,
\begin{equation} 
 \vert {\dot{\bf J}} \vert = { {K\dot{M}} \over {3 \sqrt{2} \nu_{10}} }
 (GM)^{1/2} { {C_2^{-\beta} \omega_p} \over {\beta + 1/2} } 
 \left( {\nu_{20} \over {\omega_p}} \right)^{ {2 \beta + 1} 
 \over {2 \beta + 2} },
\label{eq10} 
\end{equation}
which reduces to the formula in equation (\ref{eq7}) if $\beta = 0$. 
We have allowed ourselves a single adjustable parameter in this expression,
to best match the numerical results we adopt $C_2=0.55$.

Figure {\ref{f4}} shows how equation (\ref{eq10}) compares to the 
numerical results for choices of parameters testing the scalings implied 
by the formula. Excellent agreement (at better than the percent level) is 
obtained for two models with $\nu_1 = \nu_2$ and varying $\omega_p$, 
while a single model computed with the time-dependent code at $\beta=1.5$ also 
agrees well with the fitting formula. A model with $\nu_2 > \nu_1$ displays
some systematic error, caused by the influence of the increased ratio 
of $\nu_2 / \nu_1$ on the surface density profile in the warp region, but 
equation (\ref{eq10}) still provides a reasonable approximation to the 
numerical values for the torque. The main result is that the torque for 
a fixed accretion rate (and fixed viscous time at $R_{\rm in}$) 
drops by roughly an order of magnitude for disc 
models with $\beta$ varying between $\beta=0$ and $\beta=1.5$, while 
the dependence on other parameters ($\omega_p$, $\nu_{20}$) 
remains close to that found in the $\beta=0$ case.

\section{Alignment timescale}

\begin{figure}
 \psfig{figure=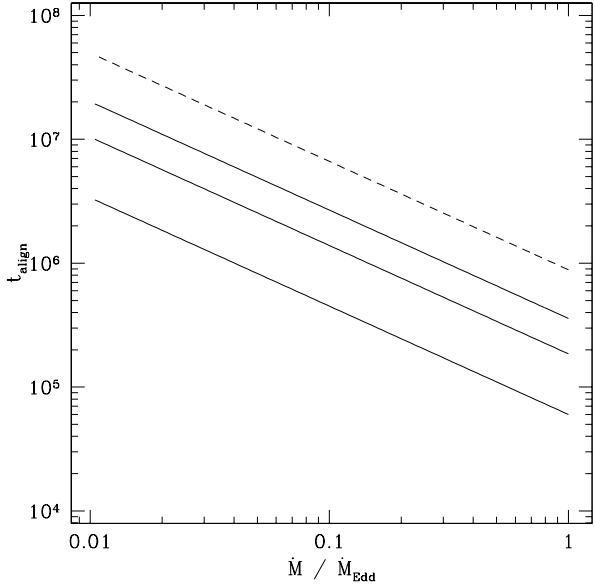,width=3.25truein,height=3.25truein}
 \caption{The timescale for the spin of a $10^8 \ M_\odot$ black hole 
          to become aligned with the angular momentum of the outer disc, plotted 
	  as a function of the mass accretion rate in units of the Eddington 
	  rate. The dashed curve is for $\nu_2 = \nu_1$, the solid curves 
	  are for $\nu_2 / \nu_1 = 1 / (2 \alpha^2)$ with (from top 
	  downwards) $\alpha = 0.3, 0.2, 0.1$.} 
 \label{f5}
\end{figure}

The torque on the black hole due to the interaction with the warped disc 
causes it to precess and become aligned with the outer disc's angular 
momentum vector on a timescale that, to factors of order unity, is just 
$t_{\rm align} = {\vert \bf{J} \vert} K / {\vert \dot{\bf{J}} \vert}$ 
(e.g. Scheuer \& Feiler 1996). ${\vert \dot{\bf{J}} \vert}$ is specified
for a given disc model by equation (\ref{eq10}), while the angular 
momentum of a hole of mass $M$ and spin parameter $a$ is 
$\vert {\bf{J}} \vert = acM(GM/c^2)$. The resulting expression for 
$t_{\rm align}$ is similar, but not equivalent, to the estimate 
given by Natarajan \& Pringle (1998), $t_{\rm align} = (| {\bf J} | / 
| {\bf J}_{\rm disk} |) (R^3_{\rm warp} / \omega_p)$, where 
${\bf J}_{\rm disk}$ is the angular momentum of the disc within a 
warp radius $R_{\rm warp}$ defined similarly to equation (\ref{eq9x}). 
Our expression takes more complete account of the need to integrate 
the torque for all radii less than $R_w$. For models of the 
disc where most of the angular momentum within the warp radius lies 
close to that radius, the final inferred timescale for alignment is
similar. 

Obtaining an estimate for $\beta$ and $\nu_{10}$ requires a model 
for the vertical structure of the disc, which serves the purpose 
of relating the unknown central disc temperature, that controls 
the viscosity, to the effective temperature fixed by the requirement 
that viscous dissipation balance radiative losses. At the radii 
of the warp, gas pressure and an electron scattering opacity 
$\kappa = 0.4 \ {\rm cm^2 g^{-1}}$ are dominant. 
In this regime, Collin-Souffrin \& Dumont (1990) obtain a radial 
profile of column density $N_{25}$ (in units of $10^{25} \ {\rm cm}^{-2}$)
described by,
\begin{eqnarray} 
 N_{25} = 98 \alpha^{-4/5} \left( {\epsilon \over 0.1} \right)^{-3/5} 
 \left( {{L / L_{\rm Edd}} \over 0.1} \right)^{2/5} \nonumber \\
 \times \left( {L \over {10^{44} \ {\rm erg/s}}} \right)^{1/5} 
 \left( R \over {10^4 R_g} \right)^{-3/5},
\label{eq11}
\end{eqnarray} 
where $\epsilon$ is the radiative efficiency, $L = \epsilon \dot{M} c^2$ 
is the bolometric luminosity, $L_{\rm Edd} = 4 \pi c^3 R_g / \kappa$ 
is the Eddington luminosity, and $R_g = 2GM / c^2$. For this model, 
$\beta = 0.6$. Although this 
disc model includes the effects of irradiation of the upper layers 
of the disc by the central source, we note that the scaling 
$\Sigma \propto \alpha^{-4/5} \dot{M}^{3/5} M^{1/5} R^{-3/5}$ is
identical to that of non-irradiated models (e.g. Sincell \& Krolik 1998). 
The normalisation is within a factor of two. Other uncertainties thus
dominate over that arising from different treatments of the vertical 
disc structure.

Figure {\ref{f5}} shows the calculated alignment timescale for a 
$10^8 \ M_\odot$ black hole accreting at rates between 
$10^{-2} \ \dot{M}_{\rm Edd}$ and $\dot{M}_{\rm Edd}$, where 
$\dot{M}_{\rm Edd} = 4 \pi c R_g / (\epsilon \kappa)$. We take 
a radiative efficiency $\epsilon = 0.1$, and assume a maximal 
Kerr hole, $a=1$. For these parameters, 
$R_w$ as defined in equation (\ref{eq9x}) is $R_w \sim 10^3 \ R_g$ for 
reasonable ratios of $\nu_2 / \nu_1$, so the main warped region 
of the disc falls self-consistently within the regime where 
gas pressure and electron scattering opacity are dominant. The 
Figure shows results calculated assuming that $\nu_2 = \nu_1$, 
along with the expectation if $\nu_2 / \nu_1 = 1 / (2 \alpha^2)$ 
for a range of plausible $\alpha$. The latter models of course 
predict more rapid realignment, as noted previously (Natarajan \& Pringle 1998).

For holes accreting at rates of the order of the Eddington limit,
the alignment timescale is found to be $10^5$ - $10^6$ yr. This 
is short compared to lifetimes of AGN, {\em irrespective} of 
what one assumes for the ratio of $\nu_2 / \nu_1$. Producing the 
giant structures observed in radio galaxies may require 
an active epoch of the order of $10^8$ yr, while some estimates for 
the quasar lifetime are similar (e.g. Haehnelt \& Rees 1993, though 
these estimates are subject to considerable uncertainty). In 
these systems we expect that there was ample time for a rotating black hole, 
whatever its initial spin axis, to become aligned with the angular 
momentum of the disc. Thus jets accelerated from the inner disc 
should be perpendicular to the plane of the outer disc, and 
constancy in the direction of such jets implies a corresponding 
stability in the disc plane over time.

\begin{figure}
 \psfig{figure=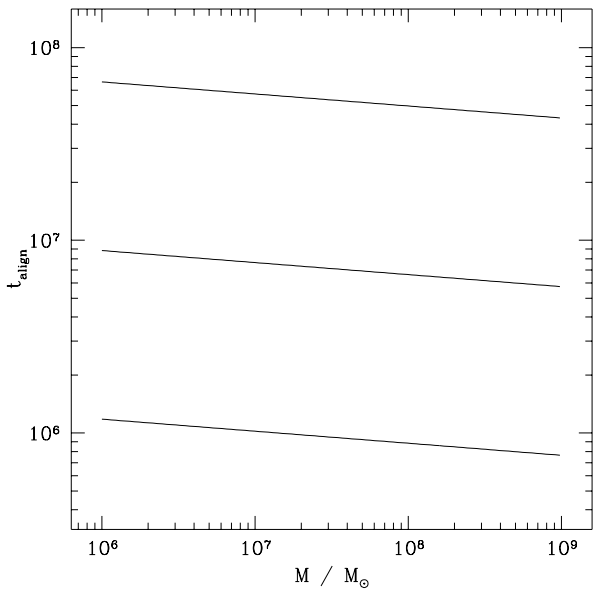,width=3.25truein,height=3.25truein}
 \caption{The timescale for the spin of the black hole to become 
 	aligned with that of the outer disc is plotted as a function 
 	of the mass $M$ of the black hole. From top downwards, 
 	the curves correspond to $\dot{M} = 10^{-2} \dot{M}_{\rm Edd}$, 
 	$\dot{M} = 10^{-1} \dot{M}_{\rm Edd}$, and 
 	$\dot{M} = \dot{M}_{\rm Edd}$ respectively.
 	We have assumed a maximally rotating 
 	hole, an $\alpha = 0.1$, and $\nu_2 = \nu_1$.} 
 \label{f6}
\end{figure}

Figure 6 plots the alignment timescale as a function of black hole mass, 
for a range of accretion rates. The alignment timescale is found to be only 
weakly dependent on the mass of the hole, as expected given the $M^{-1/16}$ 
dependence derived by Natarajan \& Pringle (1998). For this figure we 
have assumed that $\nu_1 = \nu_2$, and that $\alpha = 0.1$. Shorter 
alignment timescales are of course predicted if $\nu_2 > \nu_1$, as 
shown in Figure 5.

For holes accreting at lower rates (relative to the Eddington limit), the 
timescale for alignment grows. It is not unambiguously shorter than 
estimates for the active phase of a hole in a galactic nucleus, and 
the answer depends crucially on differences in the disc's response to 
shear and warp. It remains true that $t_{\rm align} \ll t_{\rm grow}$,
where $t_{\rm grow} = M / \dot{M}$, for any $\dot{M}$, so that 
alignment will always occur if an accretion event lasts long enough 
to contribute significantly to the hole mass. Perhaps more probably, 
however, low luminosity AGN may accrete negligible fractions of the
hole mass in many brief episodes. In this scenario, it remains possible 
that the hole spin accumulated over the whole accretion history 
will be sufficient to control the jet direction during subsequent low luminosity 
outbursts. The outer disc would then not be expected to be generally 
perpendicular to the jet direction. 

\section{Discussion}

In this paper we have considered the interaction between a misaligned 
accretion disc and a rotating black hole for the parameters appropriate 
to Active Galactic Nuclei. A combination of forced precession and disc viscosity 
allows the black hole to force the inner disc into a plane perpendicular to 
the hole's spin axis, but the resulting warped disc exerts a torque that 
eventually aligns the rotation of the hole with the angular momentum of the
outer disc. We have presented steady-state numerical solutions to the full 
warped disc equations, verified that the analytic solution of Scheuer \& 
Feiler (1996) provides an accurate representation of the disc shape and torque
on the hole, and generalised the results for a range of disc models with 
varying viscosity. The range in models we consider lead to almost an 
order of magnitude variation in the integrated torque on the hole. 

We have computed the timescale for alignment to occur, and find that for
holes accreting at rates of the order of the Eddington limit 
(for a $10^8 \ M_\odot$ 
hole, this implies $L_{\rm bolometric} = 2.5 \times 10^{46} \ {\rm erg/s}$, 
so these are luminous AGN) the timescale is short, $10^5$ - $10^6$ yr. This 
is the case for a range of assumptions as to the ratio $\nu_2 / \nu_1$, which 
characterizes the disc's relative response to azimuthal shear and to warp. 
The short derived timescale implies that the spin 
of black holes in such AGN rapidly adjusts to match the angular momentum of 
accreting gas, even if the hole gains only a small ($\ll$ 10\%) of its
mass from accretion. For low luminosity AGN the prediction is more 
model-dependent, but the timescale for alignment in these systems is
sufficiently long that the spin axis of the hole may be able to remain 
stable if individual accretion episodes are brief. Here the assumed 
value for $\nu_2 / \nu_1$ {\em is} crucial, and efforts to determine
this ratio for specific angular momentum transport mechanisms and disc 
models would be valuable.

Some powerful radio galaxies display jets that appear to have maintained 
their direction for long periods of time. Our results imply that this 
cannot be due to any intrinsic stability imparted by the spin of the black hole, 
but instead must reflect a long term constancy in the angular momentum of 
the outer regions of the accretion disc. If the gas derives from a reservoir
set up by a single accretion event then this constancy would not be 
surprising. Alternatively, a preferred axis for gas arriving in the nuclear 
regions might be the consequence of interactions between inflowing gas 
and the galactic potential.

We have calculated the torque assuming that the disc is able to relax 
into and maintain a steady-state configuration as realignment occurs. This assumption 
is justified in the strongly warped inner region, where $t_{\nu_1}$ and
$t_{\nu_2}$ are indeed smaller than the derived $t_{\rm align}$, but 
fails at large radii of $10^4$ - $10^5 \ R_g$ where our solutions show that 
the disc maintains a small but significant warp. This will not increase the 
alignment timescale (before the disc reaches a steady-state the torque in 
our time dependent calculations is {\em larger} than the limiting values
we solve for), but does suggest that following alignment of the hole the 
outer disc might retain a modest warp for a long period. For a $10^8 \ M_\odot$ 
hole a radius of $10^4 \ R_g$ corresponds to $\sim 0.1 \ {\rm pc}$, a scale which 
is observable with VLBI in nearby AGN such as NGC4258 (eg. Miyoshi et al. 
1995). Some of the warps in maser discs could then simply reflect the 
slow decay of the initial conditions of a past accretion event, and 
be devoid of a persistent forcing mechanism. Of course NGC4258 itself 
has an accretion rate much lower than the models considered here
(Gammie, Narayan \& Blandford 1999), and self-gravity is likely to 
be important in its disc (Papaloizou, Terquem \& Lin 1998), but the 
main point --- that $t_{\nu_2}$ at a few tenths of a parsec is likely 
to be long --- remains true.

A related question is whether the disc at $R \sim 10^2 \ R_g$ is able 
to attain a steady state, even in the {\em absence} of a warp. This 
region of the disc may be thermally unstable (Lin \& Shields 1986; 
Clarke 1989; Siemiginowska, Czerny \& Kostyunin 1996) and prone 
to a limit cycle involving large changes in the accretion rate. 
The interplay between such a cycle and the dynamics of a warped
disc would be complex and intriguing, though a more promising 
setting for exploring such a scenario would be in galactic 
black hole systems, if those possess warped discs. 

For AGN with rapidly spinning holes, the main warped region 
occurs at radii of $\sim 10^2 \ R_g$. During realignment 
episodes there is a substantial deposition of energy, 
originating in the spin energy of the hole, into this 
region of the disc. This is particularly the case if 
$\nu_2 \gg \nu_1$, as we have argued is likely in this 
paper and previously (Natarajan \& Pringle 1998). The
disc in the vertical structure models we have considered 
is quite thin at these radii and can easily radiate 
locally a significantly enhanced luminosity, but substantial 
changes to the vertical structure might be expected (note 
that we have assumed that the vertical structure of a warped disc
is just like that of a planar disc, which is wrong on several 
counts). Moreover, the enhanced luminosity would increase the 
disc flux at optical and infrared wavelengths, which would in 
any event be raised due solely to the greater covering fraction 
of the warped disc as seen from the central source. Since the 
alignment timescale is short, systems of this kind would be 
rare --- most AGN would harbour quite flat central discs --- but
potentially bright.

\section*{Acknowledgements}
We gratefully acknowledge useful discussions with Jim Pringle, Mitch 
Begelman, Roger Blandford, Julian Krolik and Martin Rees, and thank the 
referee for a very helpful report. PJA thanks 
Space Telescope Science Institute, where part of this paper was 
completed, for support and hospitality.

\section*{Appendix}

The ODEs for the steady state shape of the disc are most easily solved 
by expressing equation (\ref{eq1}) in terms of separate equations for the 
tilt vector ${\it \hat{l}}$ and surface density $\Sigma$. For a Keplerian 
potential these are (Pringle 1992),
\begin{eqnarray} 
 { {\partial {\bf \hat{l}}} \over {\partial t} } = 
 \left[ 3 \nu_1^\prime + { \Sigma^\prime \over \Sigma } \left(
 3 \nu_1 + {1 \over 2} \nu_2 \right) + 
 { \nu_2 \over R } \left( 
 { 3 \over 4 } + R^2 \left\vert { {\partial {\bf \hat{l}}} \over {\partial R} }
 \right\vert^2 \right) \right] { {\partial {\bf \hat{l}}} \over {\partial R} } 
 \nonumber \\
 + { \partial \over {\partial R} } \left( {1 \over 2} \nu_2 
 { {\partial {\bf \hat{l}}} \over {\partial R} } \right) + 
 {1 \over 2} \nu_2 \left\vert { {\partial {\bf \hat{l}}} \over {\partial R} }
 \right\vert^2 {\bf \hat{l}} + { {\vec \omega_p} \times {\bf \hat{l}} \over R^3},
\label{app1}
\end{eqnarray} 
where the primes denote derivatives with respect to $R$, and
\begin{eqnarray}
 { {\partial \Sigma} \over {\partial t} } &=& 
 {3 \over R} {\partial \over {\partial R}} \left[
 R^{1/2} {\partial \over {\partial R}} \left(
 \nu_1 \Sigma R^{1/2} \right) \right] \nonumber \\ &+& 
 {1 \over R} {\partial \over {\partial R}} \left[
 \nu_2 \Sigma R^2 \left\vert { {\partial {\bf \hat{l}}} \over {\partial R} }
 \right\vert^2 \right]. 
\label{app2} 
\end{eqnarray}
Setting the time derivatives to zero these constitute 3 second order equations 
for $\Sigma$ and any two components of ${\it \hat{l}}$. We solve these with 
the boundary conditions described in Section 2.3 using the NAG routines D02GAF 
and D02RAF, which implement the finite difference technique with deferred 
correction described by Pereyra (1979) (for a general introduction 
to such methods, see e.g. Press et al. 1992). An iterative approach is employed, in 
which we first solve equation (\ref{app2}) assuming zero tilt, and then use the 
resulting surface density profile in the solution of equation (\ref{app1}). The
solution for the tilt vector ${\it \hat{l}}$ is then recycled for use in 
equation (\ref{app2}), and we loop until convergence is achieved. Typically only a
small number of iterations are required, since the changes to the surface density 
profile for even quite strong warps vary smoothly with changing disc shape. A 
finely space finite difference mesh is required to obtain a solution using this 
scheme, we use between 8000 and 16,000 mesh points evenly spaced in $\log R$.
Even so, we have found that obtaining solutions with this scheme is still 
rather difficult, especially when $\beta > 0$ or $\nu_1 \neq \nu_2$. For these
cases, we start with the easy $\beta = 0$, $\nu_1 = \nu_2$ solution, and step 
towards the desired parameters using the previous solution at each step as 
the initial guess. 

We have been unable to obtain solutions using the above method for large values 
of $\beta$. In this regime, we instead evolve directly equation (\ref{eq1}) using 
the numerical code described by Pringle (1992) until a steady-state is obtained. 
This code uses an explicit first order finite difference technique, which 
conserves angular momentum to machine precision. We have modified the 
boundary conditions to be as close as possible to those described in Section 2.3, 
so that the results are directly comparable between the two solution methods. 
The time dependent code is vastly more expensive to run, for the 
runs described here we are restricted to 100 radial grid points, again spaced
evenly in $\log R$. Nevertheless this still provides reasonable accuracy for 
computing the torque.


\begin{thebibliography}{}

\bibitem{}
 Alexander P., Leahy J.P., 1987, MNRAS, 225, 1

\bibitem{}
 Bardeen J.M., Petterson J.A, 1975, ApJ, 195, L65 
 
\bibitem{}
 Blandford R.D., Znajek R.L., 1977, MNRAS, 179, 433
 
\bibitem{} 
 Cannizzo J.K., 1993, ApJ, 419, 318 
 
\bibitem{} 
 Clarke C.J., 1989, MNRAS, 235, 881 
 
\bibitem{} 
 Collin-Souffrin S., Dumont A.M., 1990, A\&A, 229, 292
 
\bibitem{} 
 Frank J., King A.R., Raine D., 1992, {\em Accretion Power 
 in Astrophysics}, Cambridge University Press (Cambridge), p.~80 
 
\bibitem{} 
 Gammie C.F., Narayan R., Blandford R., 1999, ApJ, 516, 177
 
\bibitem{}
 Ghosh, P., Abramowicz, M. A., 1997, MNRAS, 292, 887 
 
\bibitem{}
 Haehnelt M.G., Rees M.J., 1993, MNRAS, 263, 168 
 
\bibitem{} 
 Hernquist L., Mihos J.C., 1995, ApJ, 448, 41
 
\bibitem{} 
 Kumar S., Pringle J.E., 1985, MNRAS, 213, 435 
 
\bibitem{} 
 Larwood J.D., Nelson R.P., Papaloizou J.C.B., Terquem C., 1996, MNRAS, 282, 597  
 
\bibitem{}
 Lense J., Thirring H., 1918, Phys. Z., 19, 156 
 
\bibitem{} 
 Lin D.N.C., Shields G.A., 1986, ApJ, 305, 28 
 
\bibitem{}
 Liu R., Pooley G., Riley J.M., 1992, MNRAS, 257, 545 
 
\bibitem{} 
 Livio M., Ogilvie G.I., Pringle J.E., 1999, ApJ, 512, 100 

\bibitem{} 
 Maloney P.R., Begelman M.C., Nowak M.A., ApJ, 504, 77
 
\bibitem{} 
 Maloney P.R., Begelman M.C., Pringle J.E., ApJ, 474, 582
 
\bibitem{} 
 Miyoshi M., Moran J., Herrnstein J., Greenhill L., Nakai N., 
 Diamond P., Inoue M., 1995, Nature, 373, 127 
 
\bibitem{}
 Natarajan P., Pringle J.E., 1998, ApJ, 506, L97 
 
\bibitem{} 
 Ogilvie G.I., 1999, MNRAS, 304, 557
 
\bibitem{} 
 Papaloizou J.C.B., Lin D.N.C., 1995, ApJ, 438, 841  
 
\bibitem{}
 Papaloizou J.C.B., Pringle J.E., 1983, MNRAS, 202, 1181 
 
\bibitem{}
 Papaloizou J.C.B., Terquem C., Lin D.N.C., 1998, ApJ, 497, 212 
 
\bibitem{} 
 Pereyra V., 1979, in {\em Codes for Boundary Value Problems in Ordinary 
 Differential Equations}, eds. B. Childs, M. Scott, J.W. Daniel, E. Denman \& 
 P. Nelson, Springer-Verlag, Lecture Notes in Comp. Sci., 76 
 
\bibitem{}
 Press W.H., Teukolsky S.A., Vetterling W.T., Flannery B.P., 1992, 
 in {\em Numerical Recipes in FORTRAN}, Second edition, Cambridge 
 University Press, Chapter 17 
 
\bibitem{} 
 Pringle J.E., 1992, MNRAS, 258, 811  
 
\bibitem{} 
 Pringle J.E., 1996, MNRAS, 281, 357
 
\bibitem{} 
 Pringle J.E., 1997, MNRAS, 292, 136 
 
\bibitem{} 
 Pringle J.E., 1999, in {\em Astrophysical Discs}, eds. J. A.
 Sellwood and J. Goodman, ASP Conf. Ser., Vol.~160, p.~53
 
\bibitem{} 
 Rees M.J., 1978, Nature, 275, 516 
 
\bibitem{} 
 Scheuer P.A.G., 1995, MNRAS, 277, 331 

\bibitem{}
 Scheuer P.A.G., Feiler R., 1996, MNRAS, 282, 291

\bibitem{}
 Shakura N.I., Sunyaev R.A., 1973, A\&A, 24, 337
 
\bibitem{} 
 Siemiginowska A., Czerny B., 1989, MNRAS, 239, 289 
 
\bibitem{} 
 Siemiginowska A., Czerny B., Kostyunin V., 1996, ApJ, 458, 491 
 
\bibitem{} 
 Sincell M.W., Krolik J.H., 1998, ApJ, 496, 737 

\end{thebibliography}
\end{document}